\documentclass[12pt]{article}
\usepackage{times}
\usepackage{geometry}
\usepackage{aas_macros}
\usepackage[utf8]{inputenc}
\usepackage{enumitem,amssymb}
\usepackage{ragged2e}
\usepackage{natbib}
\newlist{thematic}{itemize}{8}
\setlist[thematic]{label=$\square$}

\usepackage[disable]{todonotes}
\usepackage{hyperref}

\usepackage{titlesec}
\titlespacing{\part}{0pt}{*0}{2ex}
\titlespacing{\section}{0pt}{2pt}{1ex}
\titlespacing{\subsection}{0pt}{2pt}{1ex}
\titlespacing{\subsubsection}{0pt}{*0}{1ex}

\usepackage{paralist}

\renewenvironment{thebibliography}[1]{%
\begin{oldthebibliography}{#1}%
\setlength{\parskip}{0ex}%
\setlength{\itemsep}{0ex}%
}%
{%
\end{oldthebibliography}%
}

\usepackage{pifont}
\newcommand{\swift}{\textit{Swift}}
\newcommand{\xmark}{\ding{55}}%

\newcommand{\crossbox}{\rlap{$\square$}{\large\hspace{1pt}\xmark\ }}

\begin{document}
\raggedright
\huge
Astro2020 APC White Paper 

Scheduling Discovery in the 2020s \linebreak
\normalsize

  
\noindent \textbf{Type of Activity:} \hspace{0.5cm}
\crossbox Ground Based Project \hspace{0.5cm} 
\crossbox Space Based Project \linebreak 
\crossbox Infrastructure Activity \hspace{0.5cm}
\crossbox Technological Development Activity \linebreak
$\square$ State of the Profession Consideration \hspace{0.5cm}
$\square$ Other 

\textbf{Principal Author:}

Name: Eric C. Bellm
 \linebreak						
Institution: University of Washington 
 \linebreak
Email: ecbellm@uw.edu
 \linebreak
Phone: 206-685-2112
 
\textbf{Co-authors:}
  \linebreak
Eric B. Ford (Penn State)
  \linebreak
Aaron Tohuvavohu (Penn State)
\linebreak
Michael W. Coughlin (California Institute of Technology)

\textbf{Endorsers:} 
\linebreak
Brett Morris (University of Bern)
\linebreak
Bryan Miller (Gemini Observatory)
\linebreak
Jennifer Sobeck (University of Washington)
\linebreak
Reed Riddle (California Institute of Technology)
\linebreak
Chuanfei Dong (Princeton University)
\linebreak
Peter Yoachim (University of Washington)

\section{Overview}

\textbf{The 2020s will be the most data-rich decade of astronomy in history.}  On the ground and in space, powerful facilities like LSST, JWST, massive multi-object spectroscopic surveys, and a wide variety of smaller robotic and queue-based telescopes will be repeatedly scanning the sky.  \textbf{However, without additional effort we are unlikely to realize the full scientific potential of our investments in these instruments.}

\textbf{As the scale and complexity of our surveys and instruments increase, the problem of scheduling (which observations, in what order?) becomes more critical.} First, prudence dictates maximizing the efficiency of facilities with high development costs and finite lifetimes.  Second, key scientific projects---including identifying unseen populations of compact objects, understanding stellar binarity, and discovering and classifying rare classes of transients---require complex history-dependent observational sequences.

\textbf{To date most scheduling of astronomical facilities has relied on very basic approaches}: typically manual scheduling by human operators or simple ``greedy'' algorithms with basic objective functions.  
\textbf{We argue that these approaches are insufficient for the scientific needs of the 2020s.}

Thankfully, knowledge from fields such as Operations Research is beginning to percolate into astronomy.  Surveys such as LCO, ALMA, ZTF, \swift\, and LSST  are applying new algorithms to improve their efficiency and quantitative scientific throughput.  However, much work remains to be done. 
\textbf{To maximize science in the 2020s, we must develop high-quality scheduling approaches, implement them as open-source software, and begin linking the typically separate stages of observation and data analysis.}
The latter provides real-time feedback maximizing progress towards the scientific goal--the so-called ``fifth paradigm''\footnote{The first four paradigms are observation, analytic theory, computation and simulation, and data-intensive science \citep{Hey:09:FourthParadigm}.} of science \citep{Szalay:19:FifthParadigm}. 

We provide an overview of key research directions as well as recommendations for scientists, facilities, and agencies to facilitate progress in the field.

\section{Recommendations} \label{sec:recs}
\subsection{Scientists}
\begin{compactitem}
\item Develop collaborations with scientists in operations research and related fields to help advance the state-of-the-art in astronomical scheduling and experimental design
\item Develop formalisms connecting scientific results to the observational sequences required
\item Incorporate fundamentals of experimental design into graduate curricula 
\item Continue to organize conference tracks at regular meetings like SPIE, ADASS, etc. as well as dedicated conferences\footnote{e.g., ``Artificial Intelligence in Astronomy'',
\url{https://www.eso.org/sci/meetings/2019/AIA2019.html}} to share ideas and new approaches.
\end{compactitem}

\subsection{Surveys \& Facilities}
\begin{compactitem}
\item Critically examine current scheduling and operations models and seek improvements
\item Develop or adopt automated scheduling approaches
\item Document and release scheduling tools as open-source packages to enable broad adoption
\item Document scheduling decisions, so that future statistical analyses can mitigate biases\footnote{For example, which observations of a star were part of a baseline planet survey and which observations were done in order to expedite publication (regardless of whether decision is made by computer or human)?}
\item Adopt a shared schedule reporting schema to allow for efficient coordination and contextual scheduling decisions\footnote{e.g., via the proposed IVOA Observation Locator Table Access Protocol,
 \url{http://www.ivoa.net/documents/ObsLocTAP/index.html}}
\end{compactitem}

\subsection{Funding Agencies}
\begin{compactitem}
\item Formally evaluate plans for scheduling software and experimental design as part of mission/survey proposals
\item Evaluate scheduling and operations models of current and forthcoming facilities
\item Fund interdisciplinary efforts to develop new scheduling approaches and software, particularly that with applicability beyond single missions
\item Fund contributions to general-purpose software libraries that schedulers and other high-level packages depend on.
\item Recognize contributions to open-source software as a particularly high-value form of broader impact activity.
\end{compactitem}

\section{Impact on the Field} \label{sec:impact}


In the next decade, the scientific impact of a wide range of facilities can be enhanced by improvements in scheduling.

New large facilities with queue-scheduled Guest Observer programs run by time allocation committees will merit investments in efficient scheduling approaches appropriate to their staggering budgetary scale and finite lifetimes: these include JWST and the 30\,m-class ground-based telescopes.

Large scale imaging surveys will continue to proliferate, with LSST, WFIRST, and Euclid only the most massive examples.
Effective scheduling  will be required to deliver cosmological constraints with minimal (and quantifiable) bias \citep[e.g.,][]{2019BAAS...51c.207B, 2019BAAS...51c.470C,2019BAAS...51c.418E, 2019BAAS...51c.140K}, to detect and classify rare types of transients \citep[e.g.,][]{2019BAAS...51c.339G, 2019BAAS...51c.305F}, to inventory moving objects in our solar system \citep[e.g.,][]{2019BAAS...51c.327M, 2019BAAS...51c.340C}, and to trace stellar populations in our Galaxy and beyond \citep[e.g.,][]{2019BAAS...51c.188K,2019BAAS...51c.104R, 2019BAAS...51c.206P}.  
Smaller-aperture facilities may choose to adjust their own observing strategies to complement larger facilities or to explore focused niches for scientific returns too specialized for a general purpose survey.

Likewise, other non-electromagnetic facilities such as ground-based gravitational wave interferometers (e.g., LIGO) have observational selection effects \citep{Chen_2017} that, properly exploited via coordinated scheduling, can dramatically increase the serendipitous yield of detected multi-messenger sources \citep{Tohuvavohu:Biasing}.

Some time-domain surveys such as ZTF and LSST will be issuing real-time alert streams with millions of events nightly. 
Rapid, automated prioritization and robotic followup strategies first applied to gamma-ray burst followup can now be applied to a wide spectrum of time-domain events.
Such followup will be required to fully understand rare and fast-evolving time domain events, such as kilonovae, ``orphan'' afterglows of gamma-ray bursts, rare types of stellar variables, and more.

Some of this followup will be undertaken by distributed networks of fully robotic telescopes (e.g., the Las Cumbres Observatory) which may be scheduled as a single entity \citep{Lampoudi:15:LCOGTScheduler}.
However, an even greater opportunity and challenge is to realize the potential of heterogeneous telescope networks \citep[e.g.,][]{2014htu..conf...95H} to stitch together disparate facilities in a uniform way.
This can be useful for a few purposes. 
For example, using telescopes in both hemispheres can be important due to significant size of the sky localizations associated with short gamma-ray bursts and gravitational-wave sources.
Often localizations have probability that is not fully accessible from a single location. 
It can also be useful to reimage the same localizations with multiple systems, both in different filters to measure an object's color, as well at different times to differentiate between real transients and asteroids, as well as potentially measuring a change in luminosity.
Coordinated scheduling between systems with significant differences between telescope setups, including their placement on the Earth, and their instrument configurations, including field of view, filters, typical exposure times, and limiting magnitudes, is desired. 

Existing networks have begun to implement this network-level capability in open-source codebases \citep[e.g.,][]{CoTo2018}\footnote{\url{https://github.com/mcoughlin/gwemopt}}.
The Astronomical Event Observatory Network (AEON\footnote{\url{http://ast.noao.edu/data/aeon}}) represents a current and more general step towards connecting multiple observatories through a common interface (\citealp{2018SPIE10704E..0ZS}, and see the Miller et al. 2019 APC white paper ``Infrastructure and Strategies for MMA and Time Domain Follow-Up''), with a goal of enabling network-level scheduling of both imaging and spectroscopy for multiple teams on disparate ground-based facilities around the world. 

Multi-wavelength and multi-team followup efforts, such as electromagnetic followup of gravitational wave triggers \citep[e.g.,][]{2019BAAS...51c.295F,2019BAAS...51c.276S}, present an additional layer of complexity.
Coordinating disparate and perhaps competing PI teams using key shared facilities, each with their own TOO policies and scheduling constraints, is as much a political problem as an algorithmic one.
But greater sharing of observations planned and undertaken could enable individual teams to pursue quantitatively rigorous approaches.

Massive multi-object spectroscopic surveys (DESI, PFS, SDSS-V, WEAVE, 4MOST, and more) will need to dynamically adapt their targeting strategy to account for changing observing conditions.
The SDSS-V survey in particular has a strong time-domain component which will require effective scheduling \citep[e.g.,][]{2019BAAS...51c.104R,2019BAAS...51c.503K,2019BAAS...51c.274S}.
With appropriate investment they could also opportunistically allocate spare fibers dynamically to follow up targets identified by time-domain imaging surveys.

Massive radial velocity surveys for exoplanets will demand rigorous scheduling approaches to answer key population questions \citep[e.g.,][]{2019astro2020T.466F, 2019BAAS...51c.443B}.

And finally, individual investigators with classically-scheduled nights will continue to seek to make the most effective use of the time they have available.

Across all of these diverse fields, improved scheduling methods can provide a range of benefits, including:
\begin{compactitem}
    \item Greater scientific productivity and reduced overheads
    \item Greater transparency and reproduceability 
    \item Improved ability to rigorously simulate selection functions
    \item Potentially lower operations costs if replacing labor-intensive manual scheduling approaches (e.g., manually producing several versions of queue schedules for potentially different observing condition bands, most of which won't occur).
    \item The ability to dynamically respond to failures, changing observing conditions, and target of opportunity requests by repeatedly and regularly re-solving the scheduling problem \citep[e.g.,][]{Lampoudi:15:LCOGTScheduler}.
\end{compactitem}

However, implementing state-of-the-art scheduling methods is currently not without its challenges:
\begin{compactitem}
    \item Initial development of new scheduling approaches requires specialized skills which are quite rare in astronomy.
    \item Changes to operations models and program prioritization may encounter user resistance unless the benefits (and any tradeoffs) are made clear.
\end{compactitem}

The costs of implementing such scheduling approaches would be lowered if there were widely-available open-source scheduling software\footnote{See the E.\ Tollerud et al. APC whitepaper, ``Sustaining Community-Driven Software for Astronomy in the 2020s,'' for more on the scientific importance of open-source development and recommendations for sustaining it.} for the major classes of astronomical scheduling problems.  
The need to develop such algorithms and software forms the core of our recommendations.

\section{Current Work}

Current surveys are continuing to develop innovative scheduling approaches.  These include both advances in the metrics or objective functions that surveys seek to maximize as well as new optimization algorithms.

\subsection{Objective Functions}

Many facilities on the ground and in space operate largely in a queue-based guest observer mode.
Typically these observatories have disparate instruments and/or configurations, and the observing programs are selected on a periodic basis by a time allocation committee.
Such facilities are well-served by optimization metrics that assess whether the proposals with higher TAC rank were executed in preference to those with lower rank, perhaps weighted by observing conditions.
\citet{Lampoudi:15:LCOGTScheduler} and \citet{Solar:16:SchedulingModel} describe TAC-priority objective functions for the Las Cumbres Observatory robotic telescope network and ALMA, respectively.

Given their inherent multiplexing, large-scale imaging surveys often (but not exclusively) pursue multiple science goals simultaneously. 
They are accordingly well-served by objective functions that capture the rate at which they survey spatial volume \citep[e.g.,][]{Bellm:16:Cadences,Bellm:19:ZTFScheduler} or gain information \citep{Tonry:18:ATLAS} or signal-to-noise.

Experiments that focus on one major science question can adopt metrics more specialized to their specific science goal.
Bayesian objectives are particularly suitable in these cases, as they can 
maximize the expected change in information content of the posterior for the quantity being measured \citep{AstAdaptiveSched,RvAdaptiveSched}.

\subsection{Scheduling Algorithms}

While the science metric to optimize will be specific to the survey or facility, in many cases the scheduling algorithm may be decoupled from the metric chosen.  
While astronomical scheduling problems have clear analogues in the optimization literature, most are different enough from standard textbook problems to prevent easy application of known solutions.
For example, the problem of minimizing slew time when pointing a telescope can be cast as a Travelling Salesman Problem. 
However, for ground-based telescopes the earth's rotation makes the time to slew between two points time-dependent.
Moreover, the value of observing the fields at a specific time (the science metric) also is likely to change with time, for example as the airmass of the object changes \citep[cf.][]{Bellm:19:ZTFScheduler}. 

Likewise, for space-based telescopes the slew-time between two points varies with time depending on many factors, including the exact dynamic Attitude Control System constraints of the spacecraft. Analogously, the value of observing fields at a specific time can change depending on the distance of the object from the Earth-limb at that time, for example \citep[cf.][]{Tohuvavohu_2017}.

Because many optimization problems have only heuristic solutions,
the optimization literature is large, and many approaches have been explored by different astronomical projects \citep[for a review, see][and references therein]{Solar:16:SchedulingModel}.
We highlight a handful of the wide range of approaches reported in the literature.
In many cases hybrid approaches that apply different algorithms for short- and long-term scheduling are used \citep[e.g.,][]{Colome:10:The-TJO-OAdM-Ro,Garcia-Piquer:14:CARMENES-instru,Wetter:15:CloudScheduling}.

``Greedy'' or local search-based scheduling is quite common. 
The known shortcomings of greedy searches have led some authors to propose simple extensions that take into account astronomy-specific features like targets rising and setting \citep[e.g.,][]{Denny:06:Scheduler, Rana:16:An-optimal-meth}.
Others have adopted optimization approaches that heuristically attempt to identify better solutions than the current local optimum, including simulated annealing, genetic algorithms, Tabu Search, and Ant Colony Optimization \citep[e.g.,][]{Colome:12:Research-on-sch,Garcia-Piquer:14:CARMENES-instru}.

In some cases generating a satisfactory observing sequence that avoids certain instrumental limitations is more important than maximizing an objective function.
This suggests casting the scheduling problem in Constraint Satisfaction terms.
A major example is the SPIKE system \citep{Johnston:94:HSTScheduling}, which has been used by the Hubble Space Telescope, Spitzer, Chandra, and the ground-based VLT. A related approach, TAKO (Timeline Assembler Keyword Oriented), has been utilized by \swift\ (2004--2018), \textit{Fermi}, \textit{Suzaku} and others.
The \swift\ Scheduler \citep{Tohuvavohu_2017} and the open-source \texttt{astroplan} module \citep{Morris:18:Astroplan} also employ related approaches, though the \swift\ Scheduler combines this approach with a dynamic, fuzzy weighted priority scheme \citep{Luo-2003, Verfaillie:2005} and a flexible suite of optimization (objective) functions that yield a framework with generalities and automation capabilities more similar to Integer Linear Programming approaches.

Integer Linear Programming (ILP)-based approaches have been adopted by Las Cumbres Observatory \citep{Lampoudi:15:LCOGTScheduler}, ALMA \citep{Solar:16:SchedulingModel}, and ZTF \citep{Bellm:19:ZTFScheduler}.  
These provide a more general framework for optimizing objective functions than constraint satisfaction while still enabling rigorous handling of constraints.
Powerful commercial libraries are readily available\footnote{e.g., Gurobi, IBM CPLEX}.
Casting the scheduling problem in ILP terms enables rapid, regular reoptimization to update the schedule to respond to the success or failure of observations, changing weather conditions, etc.

Finally, some schedulers are based on artificial intelligence ``agents'' that are pre-trained to make decisions about observation sequences and then run on that trained model.
Reinforcement learning approaches such as that of the LSST scheduler
\citep[e.g.,][]{Naghib:19:FeatureBasedScheduler} fall into this category.
Reinforcement learning using deep neural networks has generated a great deal of popular press thanks to Google's AlphaGo \citep{Silver:16:AlphaGo,Silver:17:AlphaGoZero,Silver:18:AlphaZero}.

This profusion of solutions highlights the potential for breakthrough new approaches, but it also raises questions.
How can we compare
the performance of disparate systems applied to different scheduling problems?
What are the performance and operational tradeoffs between different algorithmic approaches? 
Are there specific scheduling strategies that are provably superior for certain classes of problems found in astronomy?

\section{Technology Drivers} \label{sec:tech}

In addition to the exciting new astronomical facilities which motivate us to seek improved scheduling (\S \ref{sec:impact}), in the 2020s we expect to see a continued technological improvements from outside astronomy that will impact the state of the art in scheduling.

In particular, rapid progress in Artificial Intelligence algorithms driven by commercial industry will continue, particularly in the near term around deep neural networks.
Expensive optimization workflows will become more tractable due to a variety of factors. Access to scalable cloud computing will become easier and less expensive\footnote{cf.\ the A.\ Smith et al.\ APC whitepaper ``Astronomy Should be in the Clouds.''}.
Cost per CPU core should continue to decrease, aiding optimization approaches which can be parallelized.
Improved tooling for porting code to run on GPUs can yield even larger performance gains, and we are beginning to see application-specific integrated circuits such as Tensor Processing Units (TPUs) that are specifically developed for optimization workflows.

In conjunction with further algorithmic work (\S \ref{sec:plan}), these technological developments will bring currently intractable scheduling problems within reach within the next decade.
For example, we might be able to ask, ``Given all currently visible supernovae, what photometric data point should we take next to most improve our constraint on the dark energy figure of merit?''

\section{Strategic Plan} \label{sec:plan}

We prioritize five key areas for further development in the 2020s:

First, the development of a more general taxonomy of astronomical scheduling problems will illuminate mathematical commonalities and differences and guide algorithm selection.

Second, refinement of science-based metrics for optimization will enable surveys and facilities to better quantify their progress towards their scientific goals.

Third, development of more powerful scheduling algorithms will improve the overall throughput of existing facilities at modest cost.

Fourth, development of well-tested, well-documented open source implementations of scheduling algorithms will encourage reuse in the major problem domains, lowering the barrier to entry. 
The \texttt{astropy}-affiliated \texttt{astroplan} package provides one potential framework to which scheduling modules could be contributed.


Finally, connecting observations to data reduction and analysis will enable real-time optimization of a survey's ultimate science goals.
For example, an exoplanet RV survey might ask, ``How does this observation improve my knowledge of the mass of a planet?'' \citep{RvAdaptiveSched} or even better, ``How does the observation improve my knowledge of the mass distribution of 2--4 $R_\oplus$ planets with orbital periods of 30--300 d?'' 
Research in the next decade will investigate how to make this ``fifth paradigm'' of science \citep{Szalay:19:FifthParadigm} computationally feasible while preserving secondary science objectives and serendipitous discoveries.

\bibliographystyle{aasjournal}
\bibliography{main}

\end{document}